\documentclass[prl,twocolumn, superscriptaddress, showpacs,preprintnumbers,amsmath,amssymb]{revtex4}
\pdfoutput=1

\usepackage{xspace}
\usepackage{graphicx}
\usepackage{color}

\def\beq{\begin{equation}}
\def\eeq{\end{equation}}

\def\bea{\begin{eqnarray}}
\def\eea{\end{eqnarray}}

\newcommand{\roughly}[1]%
    {{\mathrel{\raise.3ex\hbox{$#1$\kern-.75em\lower1ex\hbox{$\sim$}}}}}
\newcommand{\lsim}{\mathrel{\roughly<}}

\newcommand{\al}{\ensuremath{\alpha}}
\newcommand{\be}{\ensuremath{\beta}}

\newcommand{\la}{\ensuremath{\lambda}}

\newcommand{\si}{\ensuremath{\sigma}}

\newcommand{\hc}{\ensuremath{\mbox{h.c.}}}

\newcommand{\GeV}{\ensuremath{\mathrm{~GeV}}}
\newcommand{\TeV}{\ensuremath{\mathrm{~TeV}}}

\newcommand{\Ref}[1]{Ref.~\cite{#1}}

\renewcommand{\Re}{\mathop{\rm Re}}

\definecolor{white}{rgb}{1.0,1.0,1.0}

\newcommand{\met}{\mbox{${\rm \not\! E}_{\rm T}$}}

\begin{document}


\title{Early Higgs Boson Discovery in
Non-minimal Higgs Sectors}

\author{Spencer Chang}
\email{spchang123@gmail.com}
\affiliation{%
Physics Department,~University of California Davis\\
Davis,~California 95616}%
\affiliation{Department of Physics, University of Oregon, Eugene, OR 97403}
\author{Jared A. Evans}
\email{jaredaevans@gmail.com}
\affiliation{%
Physics Department,~University of California Davis\\
Davis,~California 95616}%
\author{Markus A. Luty}%
\email{luty@physics.ucdavis.edu}
\affiliation{%
Physics Department,~University of California Davis\\
Davis,~California 95616}%


\begin{abstract}
Particle physics models with more than one Higgs boson occur
in many frameworks for physics beyond the standard model,
including supersymmetry, technicolor, composite Higgs, and ``little Higgs'' models.
If the Higgs sector contains couplings stronger than electroweak
gauge couplings, there will be heavy Higgs particles that decay
to lighter Higgs particles
plus heavy particles such as $W$, $Z$, and $t$.
This motivates searches for final states involving multiple
$W$, $Z$, $t$, and $\bar{b}b$ pairs.
A two Higgs doublet model with custodial symmetry is a useful
simplified model to describe many of these signals.
The model can be parameterized by the physical Higgs masses and
the mixing angles $\al$ and $\be$, so discovery or exclusion 
in this parameter space has a straightforward physical interpretation.
We illustrate this with a detailed analysis of
the process $gg \to A$ followed by $A \to h Z$
and $h \to WW$.
For $m_{A} \simeq 330\GeV$,
$m_{h} \simeq 200\GeV$
we can get a $4.5\si$ signal with 1~fb${}^{-1}$ of integrated
luminosity at the Large Hadron Collider.
 
\end{abstract}


\maketitle

{\it Introduction---}The Large Hadron Collider (LHC) is currently exploring
the high-energy frontier at the TeV scale, and its main
goal is to discover the origin of electroweak symmetry breaking.
Most of the effort in this direction is devoted to the search for a
Higgs boson with mass below roughly $150\GeV$.
This is motivated by the fact that the standard model with a light
Higgs boson gives a good fit to precision electroweak data,
and by supersymmetric models that predict a light Higgs boson.
However, the minimal supersymmetric model generically
predicts $m_{h} < m_Z = 91\GeV$ for the lighest Higgs mass,
violating the experimental limit $m_{h} > 114\GeV$ from searches at 
LEP \cite{Barate:2003sz}.
The Higgs mass can get loop contributions that avoid the experimental
bound, but at the cost of fine-tuning UV contributions to the Higgs mass
at the percent level.
Eliminating precisely this kind of tuning is the primary motivation
for supersymmetry, and this has motivated a great deal of work on
supersymmetric models to reduce this tuning (see for {\it e.g.\/}\ \cite{Batra:2003nj, Harnik:2003rs, Chang:2008cw} and \cite{Graham:2009gy} for a more complete list of references).
With or without supersymmetry, the precision
electroweak fit is compatible with a heavier Higgs boson if there
are additional particles with masses and couplings that break electroweak
symmetry (see for {\it e.g.\/}\ \cite{Kribs:2007nz}).

In this Letter, we study the phenomenology of non-minimal Higgs sectors
with Higgs boson masses above $2m_W$.
Non-minimal Higgs sectors occur in many frameworks for avoiding fine tuning
of the Higgs sector, including
new strong dynamics (``technicolor'') (for a review, see \cite{Hill:2002ap}),
composite Higgs (see for {\it e.g.\/}\ \cite{Kaplan:1983fs,Agashe:2004rs}),
 and ``little Higgs'' theories (for reviews, see \cite{Schmaltz:2005ky, Perelstein:2005ka}).
Heavier Higgs bosons are associated with Higgs sectors with stronger
self-couplings, and are therefore particularly well-motivated 
in  
technicolor and composite Higgs models.
In particular,
\Ref{Evans:2009ga} argued that spin-0 resonances are a generic and prominent
feature of models with strong dynamics in the electroweak symmetry
breaking sector, and the present work originated in the construction
of simplified models to describe their collider phenomenology
\cite{ttbarsimp}.
However, the collider signals we discuss are applicable to a 
much wider range of models.

Higgs bosons couplings are generally proportional to the 
masses of the particles to which they couple,
so the Higgs particles tend to decay to the heaviest particles
that are kinematically allowed.
In the present context, this implies decays to lighter Higgs bosons,
$t$, $Z$, $W$ and $b$ particles.
In particular, we can have cascade decays leading to states containing
several heavy standard model particles.

{\it Simplified Model---}%
To explore the phenomenology,
we use a 2-Higgs doublet model as a simplified model \cite{Alves:2011wf}.
This model possesses built-in, good high-energy behavior, has a simple
parameter space, and can be
simulated using state-of-the art Monte Carlo tools.
We impose $SU(2)$ custodial symmetry as well as a discrete symmetry
$H_2 \mapsto -H_2$ on the potential to simplify the model.
The Higgs potential is then
\bea
\!\!\!\!\!\!\!\!\!\!\!
V &=& m_1^2 H_1^\dagger H_1
+ m_2^2 H_2^\dagger H_2
\nonumber
\\
&& \quad 
+ \frac{\la_1}{4} (H_1^\dagger H_1)^2
+ \frac{\la_2}{4} (H_2^\dagger H_2)^2
\\
&& \quad 
+ \la_3 (H_1^\dagger H_1) (H_2^\dagger H_2)
+ \frac{\la_4}{4} (H_1^\dagger H_2 + \hc)^2.
\nonumber
\eea
The physical particles consist of the $CP$-even neutral scalars
$h, H$ and a degenerate custodial $SU(2)$ triplet
$(A, H^\pm)$.
The 6 parameters in the Higgs sector can be taken to be
$v = \sqrt{v_1^2 + v_2^2} = 246\GeV$, 
$m_{h}$, $m_{H}$, $m_{A} = m_{H^\pm}$,
and the angles $\al$ and $\be$ defined conventionally by
$\tan\be = v_1/v_2$,
$h = [ \cos \al \Re(H_1^0) - \sin \al \Re(H_2^0) ] /\sqrt{2}$.
The custodial $SU(2)$ symmetry allows us to choose $H_1$
to be the field that couples to the top quark, so all phenomenologically
important couplings of the Higgs bosons are determined by these parameters.

{\it Phenomenology---}%
We now discuss the signals beyond
those of the standard model
when some of the Higgs bosons are heavy.
If the $A$ is relatively light and has unsuppressed
coupling to the top quark, it can be produced via
$gg \to A$ (via a top loop) followed by
$A \to \bar{t}t$ or $h Z$.
If the $A$ is light but has suppressed coupling to the top quark
(large $\tan\be$) we have $gg \to H$ followed by
$H \to A Z \to h ZZ$.
If the $A$ is very heavy, the dominant signal
beyond the standard model is $gg \to H \to hh$.
All of these signals produce light Higgses, $h$, which decay either to
$\bar{b}b$ (for $m_h \lsim 2 m_W$) or $WW$/$ZZ$ (for larger $m_h$).
There are many modes, with the common feature that they involve
production of multiple heavy standard model particles.

\begin{figure}
\includegraphics[scale=0.47]{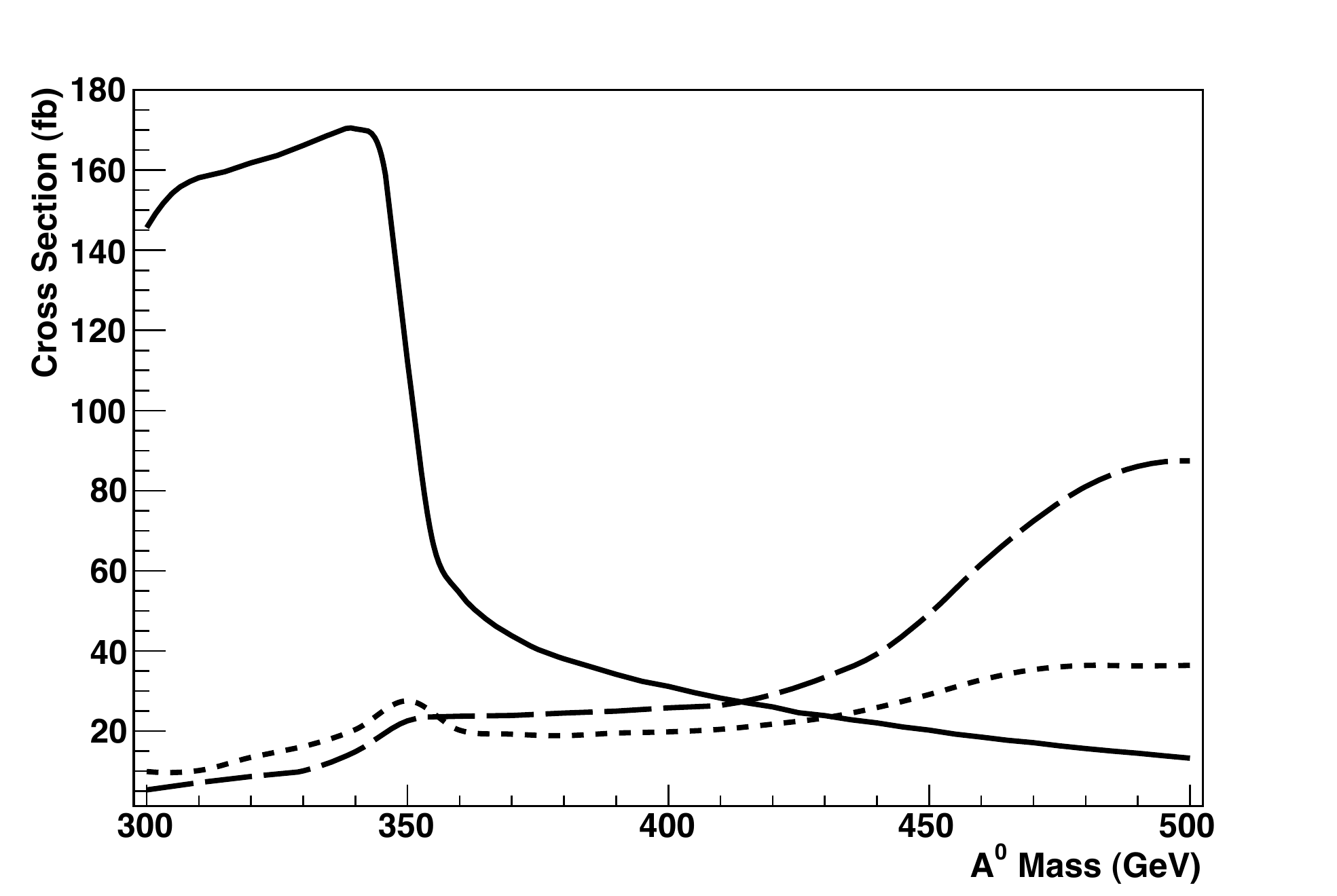}
\caption{\label{fig:A0prod}
Cross-section times branching ratios at the 7 TeV LHC
as a function of $m_A$.
Solid line: $pp \to A \to Z h \to W^+ W^- Z \to l^+ l^- l^\pm jj + \met$  
for a model with $m_h = 200\GeV$,
$\tan\be = 1$, $\sin\al = 1$.
Long dashed line: $pp\to A^* \to Z h \to b\bar{b} + \met$
with $p_T(h) > 200\GeV$ and $\met > 200\GeV$ 
for a model with $m_h = 120\GeV$,
$\tan\be = 1$, $\sin\al = 0$.
The comparable standard model cross section is 2.7~fb.
Short dashed line: $pp\to A^* \to Z h \to l^+ l^- b\bar{b}$
with $p_T(h) > 200\GeV$
for the same model.
The comparable standard model cross section is 1.1~fb.}
\end{figure}

In this work we
study the process $gg \to A^{(*)} \to Zh$.
If $m_h < 2m_W$ the dominant decay is $h \to \bar{b}b$; in the context of supersymmetry, this has been discussed as a discovery mode of two Higgs bosons \cite{Abdullin:1996as}.
$Zh$ production also occurs in the standard model,
and backgrounds can be suppressed by focusing on the kinematic region where
the $h$ is boosted \cite{Butterworth:2008iy, ATL-PHYS-PUB-2009-088}.
In the present scenario this is enhanced by the intermediate
$A$ in the production.
Requiring $p_T(h) > 200\GeV$ the intermediate $A$ is off-shell
in our model, but there is
still an enhancement of $10$ times or more compared to the
standard model, as illustrated in Fig.~\ref{fig:A0prod}.
Since this signal is already under detailed investigation
by the LHC experiments,
we will not discuss it further here.

We will study the case of a
heavier $h$, where we have $gg \to Zh \to W^+ W^- Z$
via an on-shell $A$.
To our knowledge, this
has not been previously investigated in the literature.
We focus on $Z \to l^+ l^-$ ($l = e$ or $\mu$),
which suppresses all backgrounds that do not involve
a $Z$.
We consider $WW \to jj l^\pm + \met$, which has a $29\%$ branching
ratio and allows us to reconstruct both the $h$ and the $A$.
The final state $WW \to l^+ l^- + \met$ has branching ratio
$4.5\%$ and is very clean, but does not allow reconstruction
of the $h$ (or $A$).
Another possibility that we do not investigate here
is $Zh \to ZZZ \to jj l^+ l^- l^+ l^-$,
which has a branching ratio approximately $0.2$ times
the mode we are considering.
With more data, these modes
will be useful in confirming the signal and its interpretation.

Because we require a leptonic $Z$, the only important backgrounds are
those involving a $Z$.
The largest background is $WZ + {\rm jets}$.
Other backgrounds we consider are $Z + {\rm jets}$,
$\bar{t}tZ$, and $ZZ + {\rm jets}$.
The signal was generated with up to 1 additional jet
using {\tt MadGraph} \cite{Alwall:2007st}
with the {\tt 2HDM4TC} model files \cite{ttbarsimp},
and applying a $K$-factor of $2.4$
(extrapolating from \cite{Anastasiou:2002wq}).
Backgrounds were simulated using {\tt Alpgen} \cite{Mangano:2002ea} with
up to one additional jet.
Jet showering was done by {\tt Pythia} \cite{Sjostrand:2006za}
with MLM jet matching \cite{Caravaglios:1998yr} for both the signal and backgrounds,
and the detector response was simulated using {\tt PGS} \cite{PGS}.
The $K$-factors used for the backgrounds are:
$K_{WZ} = 1.6$ \cite{Baur:1994aj,Campanario:2010xn},  $K_{Z} = 1.6$ \cite{Campbell:2002tg},  $K_{ZZ} = 1.3$ \cite{Ohnemus:1990za,Mele:1990bq},   and $K_{t\bar{t}Z} = 1.3$ \cite{Lazopoulos:2008de}.  
We study a benchmark model with
$m_A = 330\GeV$,
$m_h = 200\GeV$,
$m_H = 1\TeV$,
$\sin\al = 1$,
and $\tan\be = 1$.
The cross section for producing $A$ at the Tevatron is 190~fb.
The CDF experiment has searched for the $WWZ$ final state from 
$Zh$ associated production in the standard model \cite{TevatronHiggsWWZ}.
Although this is a neural net analysis that cannot be directly
compared with our model, the limit on the $WWZ$ cross section
is approximately 450~fb.
We conclude that Tevatron searches
are not senstive to this model, and
we focus on the prospects for discovery at the 7~TeV LHC.

For an LHC search, we
require the events to pass one of the following lepton triggers:
$(i)$ single lepton with $p_T(e) > 30\GeV$
or $p_T(\mu) > 20\GeV$;
$(ii)$ double lepton with $p_T(l_1) > 17\GeV$, $p_T(\ell_2) > 10\GeV$;
$(iii)$ triple lepton with $p_T(l_{1,2,3}) > 10\GeV$.
All triggered leptons are required to be central,
$|\eta_e| < 2.4$ and $|\eta_\mu| < 2.1$.
These triggers are similar to those used in the 2011 LHC run.
The selection cuts then require at least 3 central leptons
with $p_T(l) > 8\GeV$
and at least 2 jets with $|\eta_j| < 2.5$ and $p_T(j) > 30\GeV$.
We then impose the following cuts:
$(i)$ $\met > 20\GeV$.
This effectively suppresses the $Z + \mbox{jets}$ background.
$(ii)$ We require that 2 same-flavor, opposite-sign leptons reconstruct
to the $Z$ with $|m_{ll} - m_Z| < 7\GeV$.
We estimate that loosening this cut further will allow
$t\bar{t}$ backgrounds to be significant.
$(iii)$ We combine missing $p_T$ and the hardest remaining lepton
(and, if that fails, the next hardest remaining lepton)
to reconstruct to the $W$ mass.
This will give either 2 solutions or no solutions.
$(iv)$ Two of the jets are required to reconstruct a $W$ with
$\left| m_{jj}-m_W\right| < 25\GeV$.
The effects of these cuts are shown in Table~\ref{tab:cuts}.
Even without the cuts, the signal is visible above the background
for luminosity of order 1~fb$^{-1}$, but this requires an absolute
comparison to simulated backgrounds.
The cuts allow backgrounds to be determined by a sideband analysis,
allowing for a robust discovery or exclusion.
They are also required to reconstruct the $h$ and the $A$.
Note that these cuts do not depend on the $h$ and
$A$ masses, and should be effective for a wide range of $h$ and $A$
masses.

\begin{table}
\vspace{2mm}
\begin{tabular}{|r||c|c|c|c|c|}
 \hline
 Process & Selection & $\met$ & $Z(ll)$ & $W(l\met)$ & $W(jj)$ \\
 \hline
 \hline
  $WZ+2,3$ jets          & 8.85   & 7.94    & 7.50   &  5.70   & 1.65   \\
 $t\bar{t}Z+0,1$ jets  & 0.932 & 0.896  & 0.675 & 0.475 & 0.293  \\
 $Z+2,3$ jets              & 2.30   & 0.848  & 0.587 & 0.586 & 0.226 \\
 $ZZ+1,2,3$ jets        & 1.37   & 0.611   & 0.573 & 0.504 & 0.180  \\
 \hline
 Total  Background                         & 13.4   &  10.3   &  9.34 &   7.26 &    2.35   \\
 \hline
 Signal + \hspace{-0.7mm}$0,1$ jets    & 29.8   &  25.9   & 22.3  &  16.3 &     9.89   \\
 \hline
 \end{tabular}
\caption{\label{tab:cuts}
Signal and background cross sections (in fb) for the benchmark model
at the 7~TeV LHC.
The cuts are described in the text.
The incorrect solutions from reconstructing the leptonic $W$
are not counted in this table.}
 
\end{table}

\begin{figure}[tb]
\includegraphics[scale=0.47]{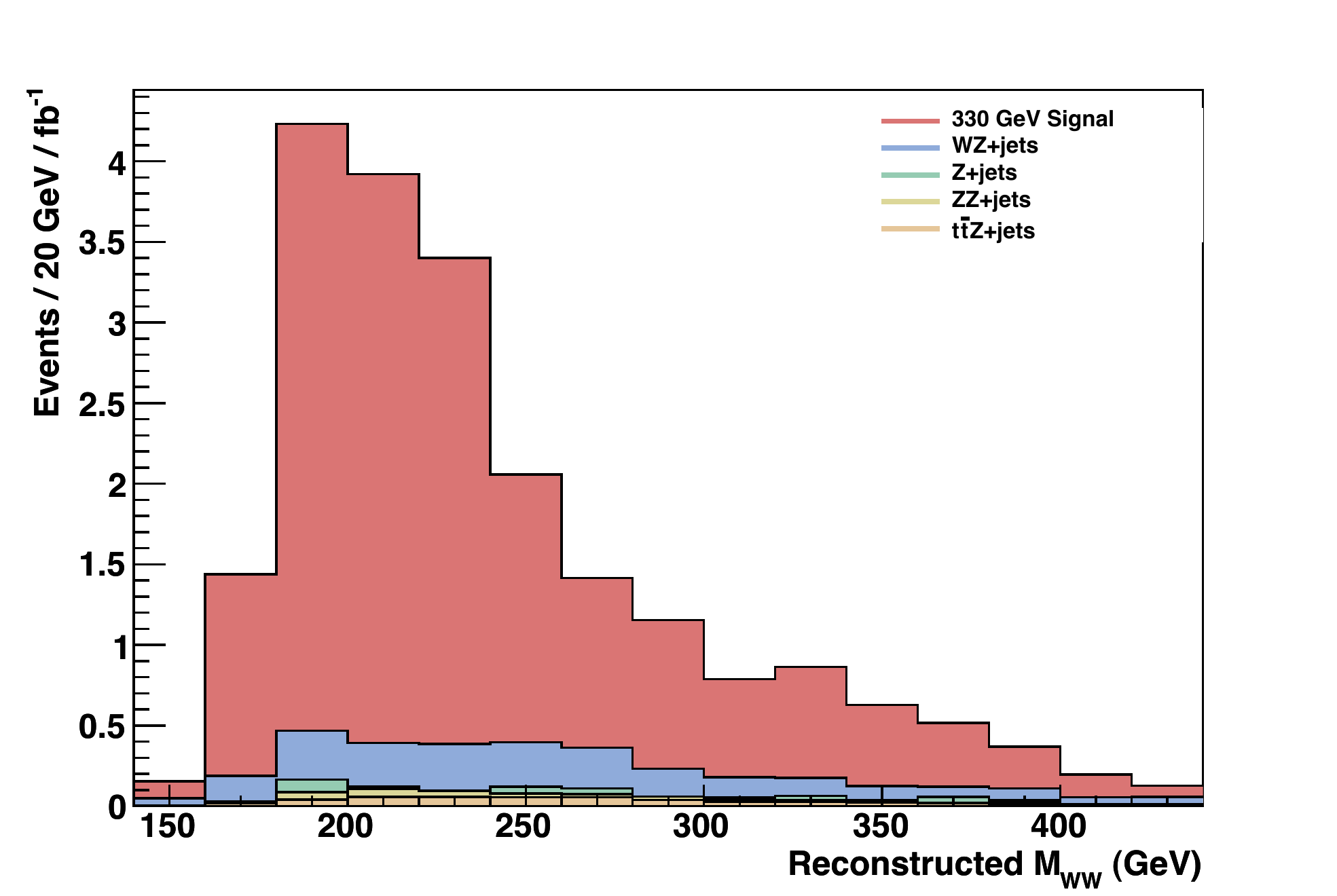}
\caption{\label{fig:MWW}
Reconstructed $h$ mass for 1~fb$^{-1}$ of data
at the 7~TeV LHC. There is a combinatoric doubling of events, since 
incorrect leptonic $W$ solutions are included.}
\end{figure}

\begin{figure}[tb]
\includegraphics[scale=0.47]{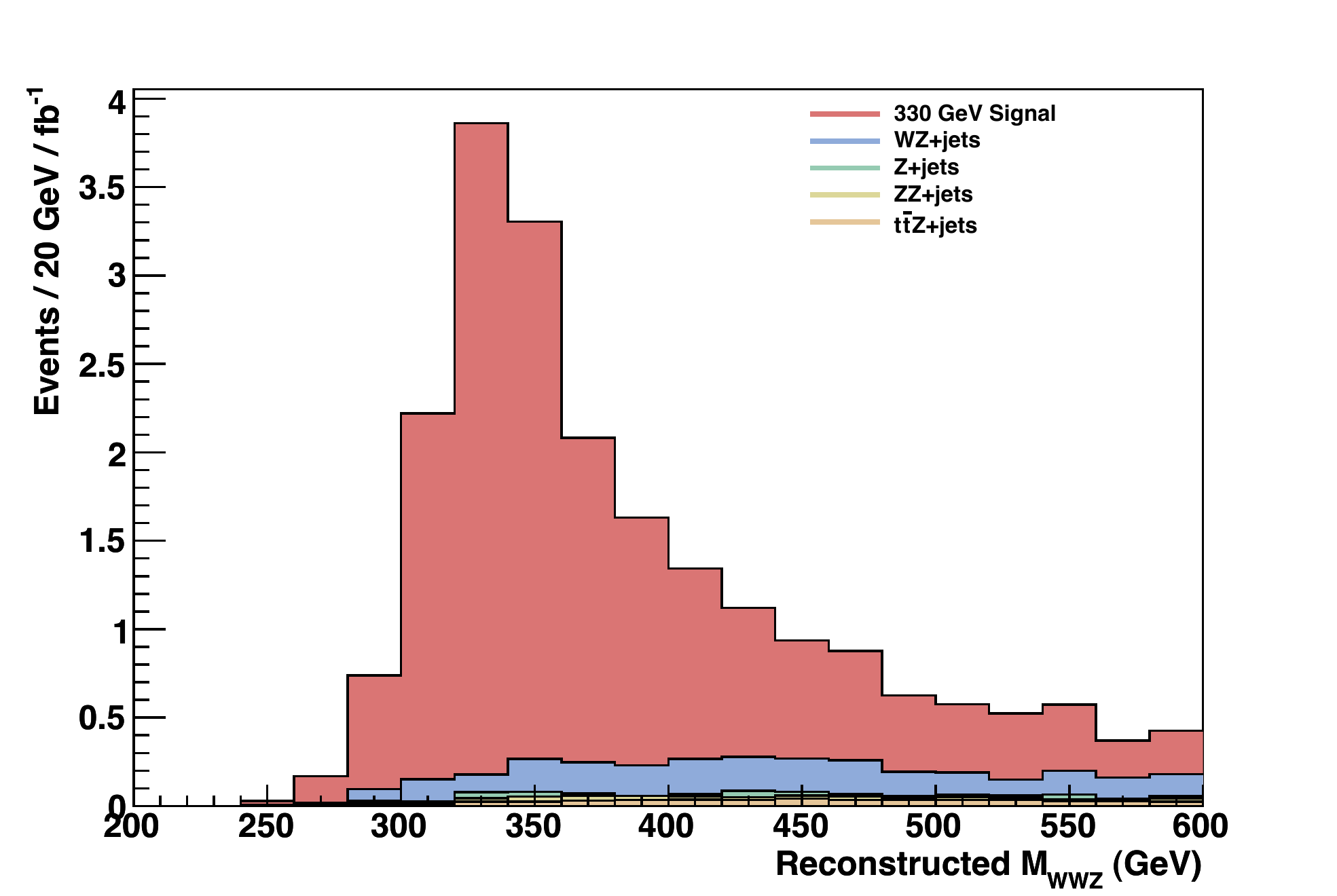}
\caption{\label{fig:MWWZ}
Reconstructed $A$ mass for 1~fb$^{-1}$ of data
at the 7~TeV LHC.  There is a combinatoric doubling of events, 
incorrect leptonic $W$ solutions are included.}
\end{figure}

After these cuts, for an LHC luminosity of
1~fb$^{-1}$ at 7 TeV running we have approximately 10 signal events,
corresponding to approximately $4.5 \sigma$ significance.
With more data, one can reconstruct the $h$ and $A$ mass peaks.
This is illustrated in Figs.~\ref{fig:MWW} and \ref{fig:MWWZ},
which are normalized to 1~fb$^{-1}$, but include additional
Monte Carlo statistics to show the shape of the peak.
These plots include the incorrect reconstructions of the leptonic $W$.
Additional techniques may
be applied to reduce the effect of these incorrect solutions,
but it is already clear that the mass can be reconstructed
from these distributions.

{\it Conclusions---}%
We have argued that models that have non-minimal Higgs sectors
with heavier Higgs bosons are well-motivated, and generically lead
to signals containing 3 or more heavy standard model
particles: $t$, $Z$, $W$, and $b$.
We have advocated a simplified 2 Higgs doublet model to
parameterize these signals.
We have illustrated the possible searches with the process
$gg \to A^{(*)} \to Z h$ followed either by
$h \to \bar{b}b$ or $WW$/$ZZ$.
For the $WWZ$ final state we demonstrated that with as little
as 1~fb$^{-1}$ of data at the 7~TeV LHC this can lead to discovery of
both the $A$ and the $h$ Higgs bosons. 
We hope that this work will motivate additional searches for
final states involving 3 or more heavy standard model particles.

\begin{acknowledgments}
{\it Acknowledgements---}%
This work was supported by DOE grant DE-FG02-91-ER40674.
We thank the UC Davis High Energy Experimental group,
particularly M.~Squires, for use of and assistance with computing.
We also thank J.~Conway, M.~Chertok, S.~Maruyama and M.~Spannowsky
for useful discussions.
\end{acknowledgments}

\bibliographystyle{apsrev}
\bibliography{WWZ}

\end{document}